\documentclass[aip, jap,preprint]{revtex4-1}
\usepackage{graphicx}%
\usepackage{dcolumn}%
\usepackage{bm}%
\usepackage{array}
\usepackage{amsfonts,amssymb, amsmath}
\usepackage{xr}
\begin{document}
\title{A kinematic study of energy barriers to crack formation in
graphene tilt boundaries}
\date{\today} 
\author{Matthew Daly}
\author{Chandra veer Singh}
\email[Corresponding author: ]{chandraveer.singh@utoronto.ca}
\affiliation{Department of Materials Science and Engineering, University of
Toronto, 184 College Street, Suite 140, Toronto, Ontario, M5S 3E4, Canada}
\begin{abstract}
Recent experimental studies have observed a surprisingly wide range of strengths in polycrystalline graphene.  Previous computational investigations of graphene tilt boundaries have highlighted the role of interfacial topology in determining mechanical properties.  However, a rigorous characterization of deformation energy barriers is lacking, which precludes direct comparison to the available experimental data. In the current study, molecular dynamics tensile studies are performed to quantify kinematic effects on failure initiation in a wide range of graphene tilt boundaries. Specifically, the process of crack formation is investigated to provide a conservative estimate of strength at experimental loading rates.
Contrary to previous studies, significant strain rate sensitivity is observed, resulting in reductions of
crack formation stresses on the order of 7 to 33\%.  Activation energies
of crack formation are calculated in the range of 0.58 to 2.07 eV based
on an Arrhenius relation that is fit to the collected simulation data.  Physically, the magnitude of activation energies in graphene tilt boundaries are found to be linearly correlated to the
pre-stress found at the critical bonds in graphene tilt boundaries. 
Predictions reported in the present study provide a possible explanation for the wide range of strengths experimentally observed in polycrystalline graphene and greatly improve upon current theoretical estimates.\\
\vspace{1cm}\\
This article has been published in {\it Journal of Applied Physics} {\bf 115} 22, 223513. doi:10.1063/1.4883190
\end{abstract}
\maketitle
\section{\label{sec:intro}Introduction}
With an intrinsic strength reported at above 100 GPa \cite{Lee2008}, graphene
permits access to previously uncharted areas of material-property space making it a desirable material for a number of composite applications\cite{Stankovich2006}. 
Efforts to increase manufacturing yield have resulted in the synthesis of
polycrystalline graphene\cite{Li2009}, with tilt boundaries separating misoriented crystallographic
domains\cite{Kim2011a}.  The impact of such tilt boundaries on mechanical properties is
currently an area of extreme research fervor, with
experimental reports of polycrystalline strength ranging from as low as 35 GPa
\cite{Ruiz-Vargas2011} to near pristine values of 98.5 GPa \cite{Lee2013}.  Weakening of polycrystalline graphene has previously been suggested as a result of high porosity in
the graphene samples\cite{Lee2013}.  A
recent
atomic force microscopy investigation of high quality graphene bicrystals,
however, has reported a wide strength envelope, with breaking stresses
encompassing approximately the entire range observed in the previous studies (48 to 83 GPa)\cite{Rasool2013}. 
The phenomena underpinning strength in polycrystalline graphene may therefore be
more complicated than sample quality considerations and merits further investigation.  

Tilt boundaries in graphene are known to be populated by topological defects\cite{An2011,Kim2011a,Huang2011a, Yakobson2011b}.  On the
atomic scale, topological defects take the form of a periodic arrangement of
heptagon-pentagon disclination dipole clusters \cite{Yazyev2010, Grantab2010,
Wei2012}.  The density and periodicity of such defect tilings is dictated by the
tessellation requirements of adjacent grains. Therefore, the tilt angle between
graphene crystals determines the spacing of disclination clusters and the
specific structure of the grain boundary.  Relative to 3D structures, planar defects such as grain boundaries
possess a greater influence on the properties of low dimensional materials.  In 2D systems, a planar defect may be considered as a flaw
transcending the entire thickness of a graphene sheet.  It is therefore expected
that the strength of polycrystalline graphene is strongly related to the
interfacial configuration connecting adjacent graphene grains.  Given the
large range of strengths reported in existing experimental studies, a rigorous study of the
strength limiting features of tilt boundaries in polycrystalline graphene is
therefore required to understand the physical phenomena underscoring weakening.

The nature of weakening in polycrystalline graphene as it relates to
tilt angle and topological structure has been examined in a number of theoretical investigations
\cite{Cao2013,Cao2012b,Cao2012,Zhang2012,Wei2012,Song2013,Kotakoski2012,
Zhang2013a,Yi2013,Grantab2010,Jhon2012}.  Notable athermal molecular dynamics
(MD) studies have identified defect-saturated high angle boundaries as
possessing both the lowest interfacial energies and the greatest strength
\cite{Grantab2010,Wei2012}, supporting recent experimental characterization
\cite{Lee2013,Rasool2013}.  The effects of temperature on the fracture behavior of graphene tilt boundaries have also been investigated briefly in theoretical studies\cite{Zhang2013a,Kotakoski2012,Song2013,Jhon2012,Zhang2012,Cao2012,
Yi2013}, with results being qualitatively comparable to the available athermal
computational investigations.  Although these numerical studies have proven
successful in establishing trends in mechanical properties, the vast majority of
these investigations are restricted to high strain rates in a relatively narrow loading range\cite{Yi2013,Cao2012b}, which may
inflate strength predictions and underestimate the impact of kinematic effects.  For instance, Yi {\it et
al.}\cite{Yi2013} performed uniaxial MD tensile simulations on a number of
graphene tilt boundaries at strain rates ranging from $\text{10}^{\text{8}}$
to $\text{10}^{\text{10}}$/s.  In this study, the authors report strain rate insensitivity with respect to strength which may be a consequence of the relatively small range of strain rates tested.  Since large interfacial
stresses have been observed in many of the sampled tilt boundary configurations \cite{Wei2012}, it is likely that kinematic effects become significant at strain rates more representative of experimental conditions (e.g. at $\sim$10$^\text{0}$/s\cite{Rasool2013}). 
 From a time-scale perspective, quantification of deformation energy barriers in graphene tilt boundaries is therefore
necessary to accurately capture strain rate sensitivity.  To the authors' knowledge, analysis of
deformation energy barriers in polycrystalline graphene is limited to a preliminary
investigation of ${\text21.7}^\circ$ tilted graphene bicrystals \cite{Cao2012b}, with strengths of 125 GPa predicted under quasi-static loading conditions.  However, current experimental reports suggest an upper limit of approximately 98.5 GPa for polycrystalline graphene\cite{Lee2013}, which indicates that kinematic effects are not fully captured by this computational study, rendering these strength predictions less accurate at experimental time-scales.  A comprehensive characterization of the deformation activation energies in graphene tilt boundaries is therefore warranted to quantify the
energy barrier resisting material fracture and inform reasonable predictions of strength.  

The purpose of the current work is to perform a comparative analysis of
kinematic effects on failure
initiation over a wide range of graphene tilt boundaries.  Specifically, the
activation energy of the initial bond-breakage event as it relates to crack
formation is selected for study.  As failure in graphene is considered to be brittle in nature\cite{Booth2008}, the crack formation stress is assumed to provide a conservative estimate of strength.  In order to capture the probabilistic nature
of the strength in polycrystalline graphene, an in-depth statistical study of critical stresses to crack formation is performed.  Both lower and
higher energy grain boundaries are studied in order to sample a wide range
of tilt angles.  Results of this study may serve to reconcile the
wide strength ranges observed in experimental testing and improve the accuracy of numerical simulations.
 
\section{\label{sec:methods}Computational methods}
MD simulations are conducted using the freely available Large-scale
Atomic/Molecular Massively Parallel Simulator (LAMMPS)\cite{Plimpton1995}.  The Adaptive
Intermolecular Reactive Empirical Bond Order (AIREBO) interatomic potential with
a bond cutoff radius of 1.92 \AA\cite{Akdim2011} is used for all MD
simulations.  In the current work, six different graphene tilt boundaries are
investigated and their relevant structural properties are summarized in Table
\ref{geo}.  For the purposes of comparison,  tilt angles reported in
previous numerical\cite{Grantab2010,Wei2012,Yi2013,Cao2012b} and experimental studies\cite{Lee2013,Rasool2013} are chosen
for investigation.  The selected tilt boundaries are characterized using coincidence
site lattice theory (CSL), following the topology construction methodology and
boundary classification system outlined in Ref. \onlinecite{Carlsson2011b}. 
Using the nomenclature of Grantab {\it et al.}\cite{Grantab2010} and Yazyev and Louie\cite{Yazyev2010}, tilt
boundaries may be further categorized into zigzag and armchair groups with the
former constructed of (1,0) and the latter with (1,0)+(0,1) disclination dipole
defects.  Disclination clusters are periodically spaced over a distance
$\text{h}_\text{d}$ as required for tessellation.  Figure \ref{simcell} provides
schematics of (1,0) and (1,0)+(0,1) disclination clusters.  For the purposes of
the current work, graphene tilt boundaries are referenced with respect to the CSL parameter,
$\Sigma$. Zigzag oriented boundaries with a tilt angle of $\theta_{\text{zz}}$ may be
described in terms of the armchair lattice angles ($\theta_{\text{ac}}$) by
the relation: $\theta_{ac} = 60 - \theta_{zz}$.  Figure \ref{topos} provides the
topologies of graphene tilt boundaries selected for study.  As shown in the figure, athermal atomic
potential energies increase in the vicinity of the disclination clusters and are
consistent with previous reports \cite{Wei2012}.  

A schematic of a typical graphene tilt boundary supercell used in MD studies is
provided in Figure \ref{simcell}.  Two anti-symmetric grain boundaries are
constructed to enforce periodic boundary conditions and avoid unwanted stress
concentration along the supercell boundaries.  Uniaxial loading is applied
perpendicular to the defect line and the simulation supercell is allowed to
contract in the longitudinal direction to accommodate Poisson effects. This loading configuration is selected in order to provide the most conservative configuration for mechanical results. Based on
previous studies, fracture is expected to originate along the disclination
cluster line \cite{Yi2013,Wei2012}, with each dipole acting as a potential
failure nucleation site.  The longitudinal dimensions of the supercell are
therefore selected to maintain a constant number of failure initiation sites
across each of the examined tilt boundaries.  All tilt boundary samples possess
at least 8000 atoms.  Strain-controlled uniaxial tensile simulations are performed using the
isothermal-isobaric (NPT) ensemble available in LAMMPS.  Prior to tensile
loading, a relaxation step is programmed to ensure a minimum system energy
and to stabilize the global temperature (T).  Stress is calculated as the spatial
and temporal average of the combined per atom virial and thermal components and
the thickness of the graphene plane is assumed to be 3.35 \AA\ \cite{Lee2008}. 
For the purposes of energy barrier calculations and strain rate sensitivity analysis, MD simulations are conducted
over a large span of loading rates.  Tensile studies are undertaken in the range
of 5x10$^\text{6}$ to 10$^\text{9}$/s, which provides a much wider sampling of kinematic effects
than previously accessed by computational studies of graphene tilt boundaries.  Statistical replication of each simulation condition is achieved using the
Gaussian random seed method and all simulations are conducted with a timestep of
1 fs.  Visualization of atomic topologies is achieved using the AtomEye atomistic configuration viewer\cite{Li2003a}.

\section{\label{subsec:energy}Energy barrier analysis of crack formation}
Energy barriers resisting failure initiation may be quantitatively
sampled through measurement of global loading conditions at the instant of
bond-breakage and subsequent crack formation.  Since graphene is known to exhibit brittle fracture\cite{Booth2008}, taking the crack formation event as strength limiting permits a conservative estimate for the bounds on graphene strength. In all MD simulations, material failure is observed to initiate along the defect line at a disclination cluster.  Figure \ref{crack}
presents typical topologies in zigzag and armchair oriented tilt boundaries at
the instant of crack formation.  Crack formation may be identified by monitoring the atomic coordination of critical bonds.  In most simulations, cracks nucleate along the
bond shared between the heptagon-hexagon carbon rings.  In some simulations of
zigzag oriented tilt boundaries, however, failure initiates from the
heptagon-pentagon bond.  Given the relatively small statistical scatter in each
tested simulation condition, the kinematics surrounding these deformation events
are expected to be quantitatively similar.  

Using the thermal activation theory of Eyring\cite{Halsey1945}, the Arrhenius
relationship may be used to describe the lifetime $\tau$ of a specimen as a
function of loading $\sigma$ and temperature T by the relation:
\begin{equation}
\tau = \frac{\tau_o}{n_s}exp\left(\frac{E_o - V_a\sigma}{k_bT}\right) \label{arhenius}	
\end{equation}
where $\tau_\text{o}$ is related to the vibrational frequency of crystalline
oscillations, $\text{n}_\text{s}$ is the number of sites available for thermal
activation, $\text{E}_\text{o}$ is the athermal activation energy of the
deformation event, $\text{V}_\text{a}$ is the activation volume, and
$\text{k}_\text{b}$ is the Boltzmann constant.  Following the analytical
formulation provided by Zhao and Aluru\cite{Zhao2010}, Eq. (\ref{arhenius}) may
be used in combination with Bailey's principle\cite{Bailey1939} to provide
unique equations for the expectation time ($t_{\text{c}}$) and stress of crack
formation ($\sigma_\text{c}$) as functions of $\dot\epsilon$ and T.  In order to reconcile strain
rate sensitivity with this analytical approach, an additional constitutive
relation for the time dependent applied stress (i.e. $\sigma =
\sigma(\text{t})$) is required.  The non-linear elastic response of graphene may
be accurately represented by a logarithmic function of the form:
\begin{equation}
\sigma(t) = a\ln (b\dot\epsilon t + 1) 	\label{modulus}
\end{equation}

\begin{figure}[t]
\centering
\includegraphics[width=1.0\textwidth, keepaspectratio]{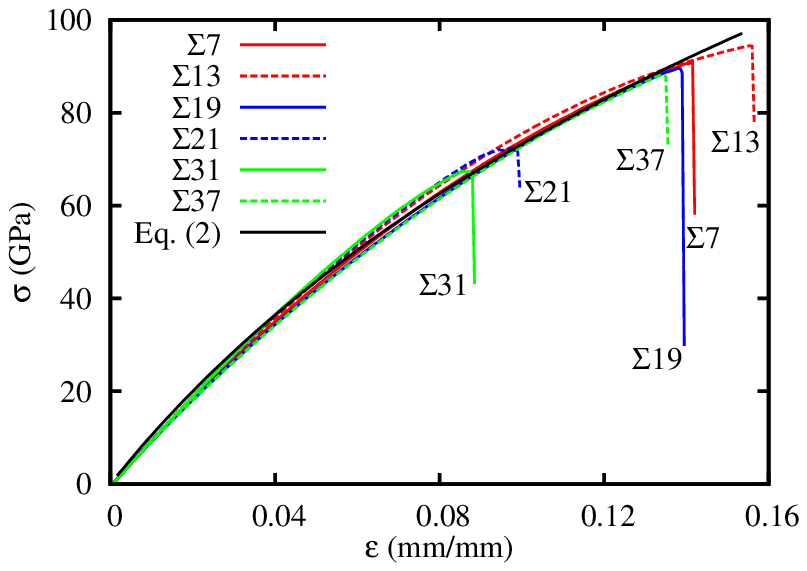}
\caption{Representative tensile simulations for each of the sampled tilt
boundaries.  Eq. (\ref{modulus}) is overlaid to show the fitted non-linear
elastic response that is assumed for energy barrier calculations.  MD
tensile simulations are conducted here at T = 300 K and $\dot\epsilon$ =
10$^{\text{9}}$/s.}
\label{tensile}
\end{figure}

Representative MD tensile simulations of each graphene tilt boundary are provided in Figure \ref{tensile}. All tilt boundaries exhibit a similar mechanical response and therefore only a singular form of Eq. (\ref{modulus}) is required to capture the non-linear elastic behavior of each grain boundary.  A least squares fit to the collected data yields a = 93.25 GPa and b = 11.94.  Eq. (\ref{modulus}) can be shown to reduce to a linear relation of
$\sigma$ $\approx$ $ab\epsilon$ \cite{Zhao2010}, where $ab$ = 1.11 TPa, which is
approximately equal to the experimentally measured in-plane modulus of 1.02 TPa
\cite{Lee2008}.  If t = t$_\text{c}$ then the crack formation stress may be defined as $\sigma_c = a\ln (b\dot\epsilon t_c
+1)$.  As per Ref. \onlinecite{Zhao2010} substitution of Eq. (\ref{modulus}) and
(\ref{arhenius}) into the Bailey criterion with t = t$_\text{c}$ provides a unique expression for
$\sigma_c$ of the form:
\begin{equation}
\sigma_c(\dot\epsilon,T) =
\frac{ak_bT}{V_aa+k_bT}\left\{\frac{E_o}{k_bT}+\ln\left[\frac{
b\dot\epsilon\tau_o}{n_s}\left(\frac{V_aa}{k_bT}+1\right)\right]\right\}
\label{strainsens}
\end{equation}
The expectation time of crack formation may then be defined as:
% \begin{equation}
% t_c = \frac{1}{b\dot\epsilon}\left[\exp\left(\frac{\sigma_c}{a}\right) -1\right]
% \label{expectation}
% \end{equation}
\begin{equation}
 t_c = \tau\left(\frac{V_a}{k_bT}+1\right)\left(1-\frac{1}{e^{\frac{\sigma_c}{a}}}\right)
 \label{expectation}
\end{equation}

\begin{figure}[t]
\centering
\includegraphics[width=1.0\textwidth, keepaspectratio]{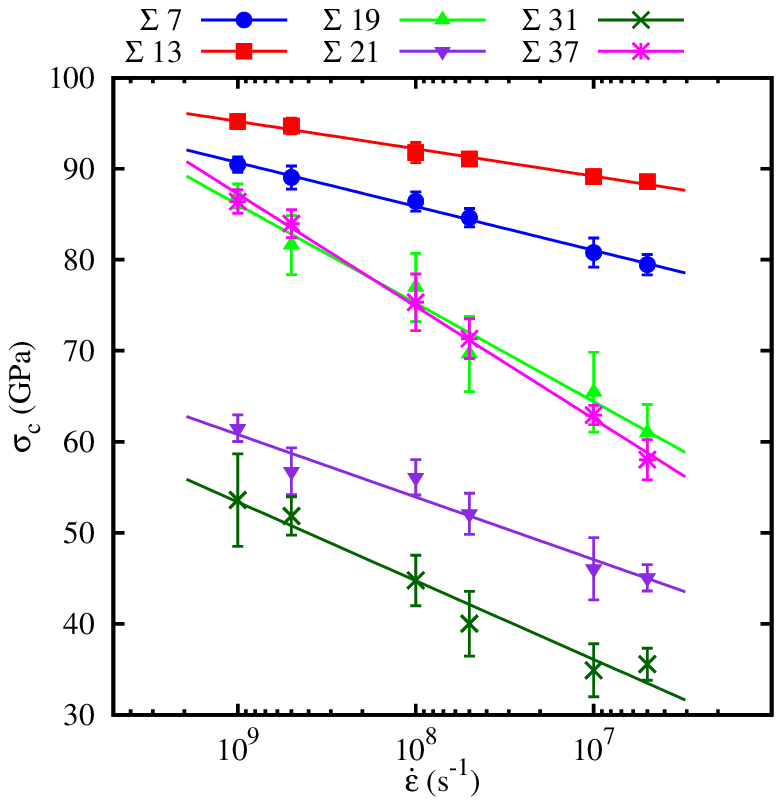}
\caption{Crack formation stresses $\sigma_\text{c}$ of the graphene tilt boundaries as determined
by MD study (T = 300 K).  All tilt boundaries are observed to exhibit some degree of strain rate sensitivity.  Eq. (\ref{strainsens}) is calculated based on fitting
of activation energies and volumes and is plotted for each tilt boundary in the respective colored stroke, showing good agreement
with MD data.  Error bars represent 95\% confidence (n = 10).  In some cases
error bars fall inside the perimeter of the data markers.}
\label{sens}
\end{figure}

Application of Eqs. (\ref{strainsens}) and (\ref{expectation}) permits direct
analysis of the energy barrier to crack formation.  However, before the
presented Arrhenius formulation is applied to the tilted graphene
samples,  MD simulation results should be validated against existing experimental data to
provide confidence in methodology.  Energy barrier analysis of pristine graphene
using the experimentally determined in-plane bond dissociation energy (4.93 eV
\cite{Brenner2002}) of graphite results in excellent agreement with MD data. 
Further details of the kinematic study of pristine graphene are
provided as supplementary data\cite{supp}.  The collected crack
formation stresses for the graphene tilt boundary samples at T = 300 K are
provided in Figure \ref{sens}, with error bars representing 95\% confidence.  Each of the sampled tilt boundaries are found to exhibit some
degree of strain rate sensitivity.  The smallest reductions in crack
formation stress are observed in the $\Sigma$ 13 tilt boundary whereas the
largest reductions occur in the 
$\Sigma$ 31 samples.  These extrema of strain rate sensitivity represent
reductions ranging from 7\% to 33\% over the approximately four orders of magnitude
of sampled loading rates.  These results suggest that in some cases
graphene can exhibit a significant degree of strain rate sensitivity and seem to contradict
 previous reports declaring the insensitivity of polycrystalline graphene\cite{Yi2013}. % Furthermore, the presented results inform more
% reasonable bounds of the quasi-static intrinsic strength of tilt
% boundaries in graphene, which is most recently reported in numerical studies as
% 125 GPa\cite{Cao2012b} ($\Sigma$ 7 boundary)  but in experimental studies as in
% the range of 48 - 83 GPa \cite{Rasool2013}.  It should be noted that the existing
% offering of numerical strain rate sensitivity studies sample strain rates over
% only 1-2 orders 
% of magnitude which may ultimately limit the applicability of their results.

\begin{figure}[t]
\centering
\includegraphics[width=1.0\textwidth, keepaspectratio]{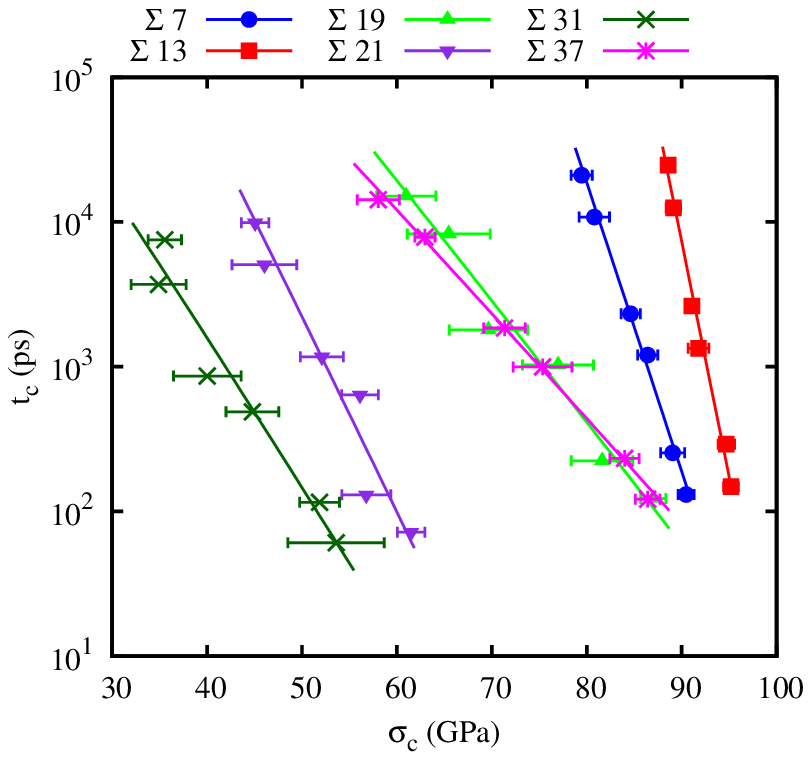}
\caption{Expectation times t$_\text{c}$ for crack formation as determined by MD study (T =
300K).  Eq. (\ref{expectation}) is calculated using activation energies and
volumes fit from Eq. (\ref{strainsens}) and overlaid for each tilt boundary in the respective colored stroke.  Error bars represent 95\% confidence (n = 10).  In some cases
error bars fall inside the perimeter of the data markers.}
\label{exptime}
\end{figure}

The collected MD data presented in Figure \ref{sens} is fit to Eq. (\ref{strainsens}) in order to determine the activation energies of crack
formation.  The fitted activation energies and volumes are then applied to Eq.
(\ref{strainsens}) and overlaid with the collected crack formation stresses,
showing excellent correlation with the MD data.  In all fitting of tilt boundary
data $\text{n}_\text{s}$ = 100, and $\tau_\text{o}$ = 0.1 ps\cite{Zhurkov1965}. 
Activation energies are found to follow a similar trend to strain rate
sensitivities and range from 0.58 ($\Sigma$ 31) to 2.07 eV ($\Sigma$ 13).  Activation volumes are found to be in the range
of $\sim$1-3\AA$^\text{3}$ which is approximately the size of a sp$^\text{2}$
covalent bond.  The expectation time $\text{t}_\text{c}$ may be also evaluated
using Eq. (\ref{expectation}) from the activation energies obtained from Eq.
(\ref{strainsens}).  Figure \ref{exptime} provides the predicted expectation
times for crack formation as a function of critical stress for each tilt boundary.  Agreement
with the collected MD data is good, providing confidence in the activation
energies obtained from Eq. (\ref{strainsens}) and the validity of the presented
energy barrier formulation.   In order to validate the robustness of the
analytical formulation with respect to temperature variations, a parallel energy barrier
analysis of the $\Sigma$ 7 boundary at T = 450 K is performed using
$\text{E}_\text{o}$ and $\text{V}_\text{a}$ obtained from fitting the data in Figure \ref{sens}. 
The results of this comparative analysis show excellent agreement across both thermal conditions and are provided in the supplementary material\cite{supp}. 

\begin{figure}[!h]
\centering
\includegraphics[width=0.8\textwidth, keepaspectratio]{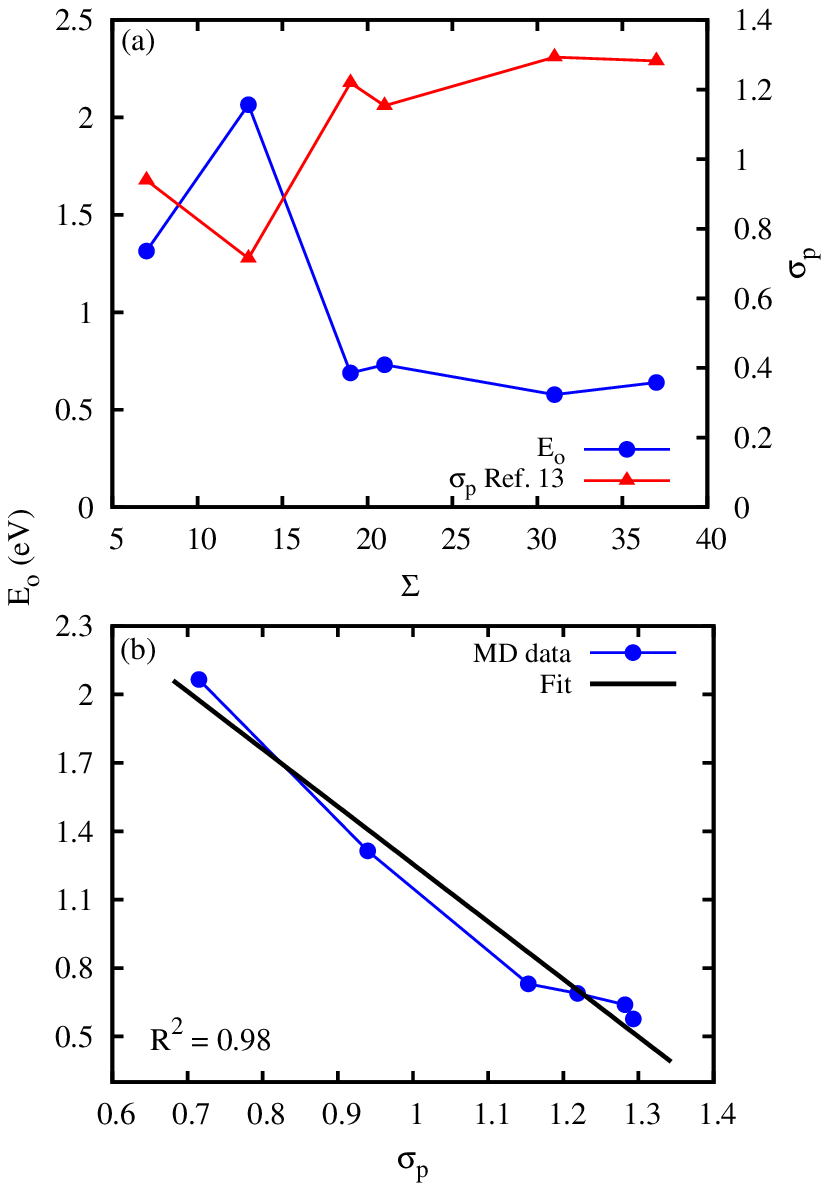}
\caption{(a) Activation energies E$_\text{o}$ are plotted alongside the normalized disclination pre-stress $\sigma_\text{p}$ in the critical bond to crack formation.  The pre-stress data is obtained from Ref.
\onlinecite{Wei2012}.  (b) Correlation of activation energy to bond pre-stress
showing a strong linear relationship (R$^\text{2}$ = 0.98).}
\label{act}
\end{figure}

In order to establish a physical rationale for trends in the calculated
activation energies, the pre-stress arising from tessellation mismatches in the
interfacial structures of the sampled tilted boundaries is considered. 
Examination of relaxed interfacial topologies shows that pre-stress in bonds
range from -75 to 90 GPa, in the $\Sigma$ 13 and $\Sigma$ 31 tilt boundaries,
respectively, with the critical bond in the $\Sigma$ 31 tilt boundary loaded to
near the upper tensile limit of the colormap (Figure \ref{topopre}).  These
large tensile and compressive stresses in the critical crack-forming bonds are
responsible for the observed differences in activation energies and premature
cracking in higher energy tilt boundaries.  The pre-stress
arising from the periodic tiling of disclination clusters in graphene tilt
boundaries has been studied in depth by Wei et al.\cite{Wei2012} and is reported here for comparison. 
Figure \ref{act} provides the computed activation energies with the disclination
normalized  pre-stress ($\sigma_\text{p}$) in the critical carbon-carbon bond of
each tilt boundary.  Examination of the
plotted data shows a remarkable correlation (R$^\text{2}$ = 0.98) between activation energy and bond
pre-stress.  Furthermore, if the
linear relation is extrapolated to $\sigma_\text{p}$ = 0, a value of 4.74 eV is
predicted, which deviates by only 4\% from the expected bond dissociation
energy of pristine graphene (4.93 eV\cite{Brenner2002}).  This finding shows that
the bond pre-stress arising from interfacial structure has a critical role in
determining the activation energies of crack formation processes in graphene. 
Analytically, the bond pre-stress serves to lower the energy barrier
($\text{E}_\text{o}\ -\ \text{V}_\text{a}\sigma$) from Eq. (\ref{arhenius}). 
Graphene samples with a low energy tilt boundary (e.g. $\Sigma$ 13) are
therefore expected to have larger crack formation stresses and thus higher
strength, whereas higher energy interfacial topologies are more prone to crack
formation.  

Extrapolation of the collected fitting results shows that kinematic effects become more pronounced as strain rates are reduced.  For
example, at a strain rate of 10$^{\text{9}}$/s and T = 300 K, MD results
predict crack formation strengths of  90.5 and 95.2 GPa for $\Sigma$ 7 and
$\Sigma$ 13 tilt boundaries, respectively.  However, by extending Eq. (\ref{strainsens}) to
strain rates typical of experimental indentation studies (e.g. 10$^\text{0}$/s)
 crack formation stresses of 47.3 ($\Sigma$ 7) and 68.1 GPa ($\Sigma$ 13) are
predicted.  A similar calculation performed on pristine graphene loaded in the armchair direction yields a crack formation stress of 84.2 GPa.  These predictions thus approximately span the range of strengths experimentally measured by Rasool {\it et al.}\cite{Rasool2013} (48 to 83 GPa) and fall between the bounds reported by Ruiz-Vargas {\it et al.}\cite{Ruiz-Vargas2011} (35 GPa) and Lee {\it et al.}\cite{Lee2013} (98.5 GPa).  Additionally, the results highlight the sensitivity of polycrystalline graphene to interfacial topology and provide a physical interpretation for the degree of weakening observed in experimental reports.  Nonetheless, caution must be exercised when making direct comparisons to experiments as the precise topology of the indented tilt boundaries is unknown. Even so, the predictions presented here greatly improve on the existing theoretical estimates (e.g. 125 GPa in Ref. \onlinecite{Cao2012b}).  The implication of these predictions is that the calculated activation energies may be used 
to estimate a conservative range for strengths in polycrystalline graphene.  The current analysis also forecasts that in some tilt boundaries (e.g. $\Sigma$ 31) bond-breakage may occur spontaneously given sufficiently low loading rates and
high enough temperatures.  This result may be rationalized by considering the
large tensile pre-stresses found in some interfacial topologies (Figure
\ref{topopre}), but requires confirmation with experimental observations.

%   As shown in this study, the effects of thermal
% activation significantly lower the strength of some graphene tilt boundaries to a much greater
% degree than previously reported, creating a wide envelope of boundary-dependent material properties.  It is therefore
% expected that the range of strengths reported in Ref. \onlinecite{Rasool2013} (48 - 83 GPa) as
% well as the upper and lower estimates of strength reported in Refs.
% \onlinecite{Lee2013} (98.5 GPa) and \onlinecite{Ruiz-Vargas2011} (35 GPa) result from
% differences in relative tilt angle and topological structure between grain
% boundaries, contrary to speculations of poor sample quality
% \cite{Lee2013}.

\section{Conclusions}
The impact of kinematic effects on the crack formation stress of graphene tilt
boundaries were studied via MD simulation.  Results of uniaxial
tensile tests indicated that, contrary to previous studies, some tilt boundaries
in graphene exhibit a large degree of strain rate sensitivity.  Higher energy tilt boundaries such as the $\Sigma$ 31 were found to be the most sensitive to loading rate, whereas lower energy boundaries such as $\Sigma$ 13 were less sensitive.  Based on MD data, an Arrhenius relationship was fit to tensile results to obtain activation energies for crack formation in the examined grain boundary configurations. 
The resultant activation energies were shown to correlate strongly to the degree
of pre-stress in the critical interfacial bonds for each topological structure. 
Although most graphene tilt boundaries showed high strength at the 
relatively high strain rates applied in MD simulations, kinematic effects were found to become more pronounced when loading
rates approached experimental ranges, leading to a considerable drop in crack
formation stresses.  In comparison to existing numerical studies, the conservative predictions of strength reported in the current study were found to be much closer to experimental observations. The range of activation energies calculated in this study highlights the importance of interfacial
topology in determining the mechanical properties of graphene tilt
boundaries and serves to rationalize the wide spectrum of experimentally reported strengths for
polycrystalline graphene.

\begin{acknowledgments}
 The authors would like to acknowledge Dr. Tobin Filleter for useful discussions
 and the Natural Sciences and Engineering Research
Council of Canada (NSERC) for providing funding for this work. Computations were performed on the GPC supercomputer at the SciNet HPC Consortium\cite{Loken2010} and the Briaree computing cluster under the administration of Calculquebec. SciNet is funded by: the Canada Foundation for Innovation under the auspices of Compute Canada; the Government of Ontario; Ontario Research Fund - Research Excellence; and the University of Toronto.
\end{acknowledgments}
\clearpage
\begin{figure}[t]
\centering
\caption{(a) A representative ($\Sigma$ 7) simulation supercell
used in MD simulations.  The loading direction is indicated in the figure and
the colormap represents atomic potential energy at 0 K. Schematics of the disclination clusters
which form the interfacial structure of (1,0) zigzag (b) and (1,0)+(0,1) armchair (c) graphene
tilt boundaries.  The distance between periodic images, $\text{h}_\text{d}$, is indicated
in each illustration.}
\label{simcell}
\end{figure}

\begin{figure}[t]
\centering
\caption{Interfacial topologies of the $\Sigma$ 7 (a),
$\Sigma$ 19 (b), $\Sigma$ 37 (c) zigzag; and $\Sigma$ 13 (d), $\Sigma$ 21 (e), $\Sigma$ 31 armchair graphene tilt boundaries.  Disclination clusters are
outlined in the black stroke and colormap is the same as in Figure
\ref{simcell}.}
\label{topos}
\end{figure}

% \begin{figure}[t]
% \centering
% \includegraphics[width=1.0\textwidth, keepaspectratio]{tensile_combined.eps}
% \caption{Collected ultimate tensile strengths $\sigma_{\text{UTS}}$ for the graphene
% tilt boundaries across different loading rates and testing temperatures.  Error bars are reported in terms of 95\% confidence with ten
% repetitions (n = 10) at each testing condition.  In many cases error bars fall
% inside the marker perimeter.}
% \label{stre}
% \end{figure}
% 
% \begin{figure}[t]
% \centering
% \includegraphics[width=1.0\textwidth, keepaspectratio]{flaw_stress.eps}
% \caption{Stress state of graphene tilt boundaries at fracture and crack
% formation.  The flaw sensitivity of tilt boundaries is visibly related to the
% disclination cluster density h$_\text{d}$. MD testing conditions are T = 300 K and
% $\dot\epsilon$ = 10$^{\text{7}}$ /s.  Error bars represent 95\% confidence (n =
% 10).  In some cases error bars are obscured by data markers.}
% \label{flaw}
% \end{figure}

\begin{figure}[t]
\centering
\caption{Atomic topologies of $\Sigma$ 7 (a), (b) and $\Sigma$ 31 (c), (d) tilt
boundaries immediately prior to; and after crack formation at 300 K.  The crack formation stress $\sigma_\text{c}$ is defined as the global stress state at the instant of crack initiation.  The
heptagon-hexagon bond is typically found to be the critical bond in crack formation.  The
inset indicates atomic coordination with black and white representing
coordinations of 2 and 3, respectively.  The colormap represents per atom stress
values along the loading direction.  Disclination clusters are highlighted in
black stroke.}
\label{crack}
\end{figure}

\begin{figure}[t]
\centering
\caption{Atomic topologies of the relaxed $\Sigma$ 13 (a) and $\Sigma$ 31 (b)
tilt boundaries at T = 0 K.  Pre-stress at the critical heptagon-hexagon bond
reaches nearly 90 GPa in the $\Sigma$ 31 structure.  Colormap indicates per atom
stresses along the loading direction and disclination clusters are outlined in
black stroke.}
\label{topopre}
\end{figure}

\clearpage
\begin{table}[!h]
\caption{\label{geo}Geometric parameters used to construct and classify the
tilt boundaries studied in MD simulations.}
\begin{ruledtabular}
\begin{tabular}{llll}
Orientation & $\theta_{\text{zz}}(^\circ)$\footnote{$\theta_{\text{ac}}$ is
provided in brackets where applicable.} & $\text{h}_\text{d}$(\AA) & $\Sigma$\\
zigzag & 21.8 & 6.507 & 7\\
zigzag & 13.2 & 10.721 & 19\\
zigzag & 9.5 & 14.961 & 37\\
armchair & 32.2 (27.8) & 8.868 & 13\\
armchair & 38.2 (21.8) & 11.271 & 21\\
armchair & 42.1 (17.9) & 13.693 & 31\\
\end{tabular}
\end{ruledtabular}
\end{table}
\end{document}

\maketitle
\section*{Supplementary Materials}

\subsection*{Energy barrier analysis of pristine graphene}
\begin{figure}[b]
\centering
\includegraphics[width=0.8\textwidth, keepaspectratio]{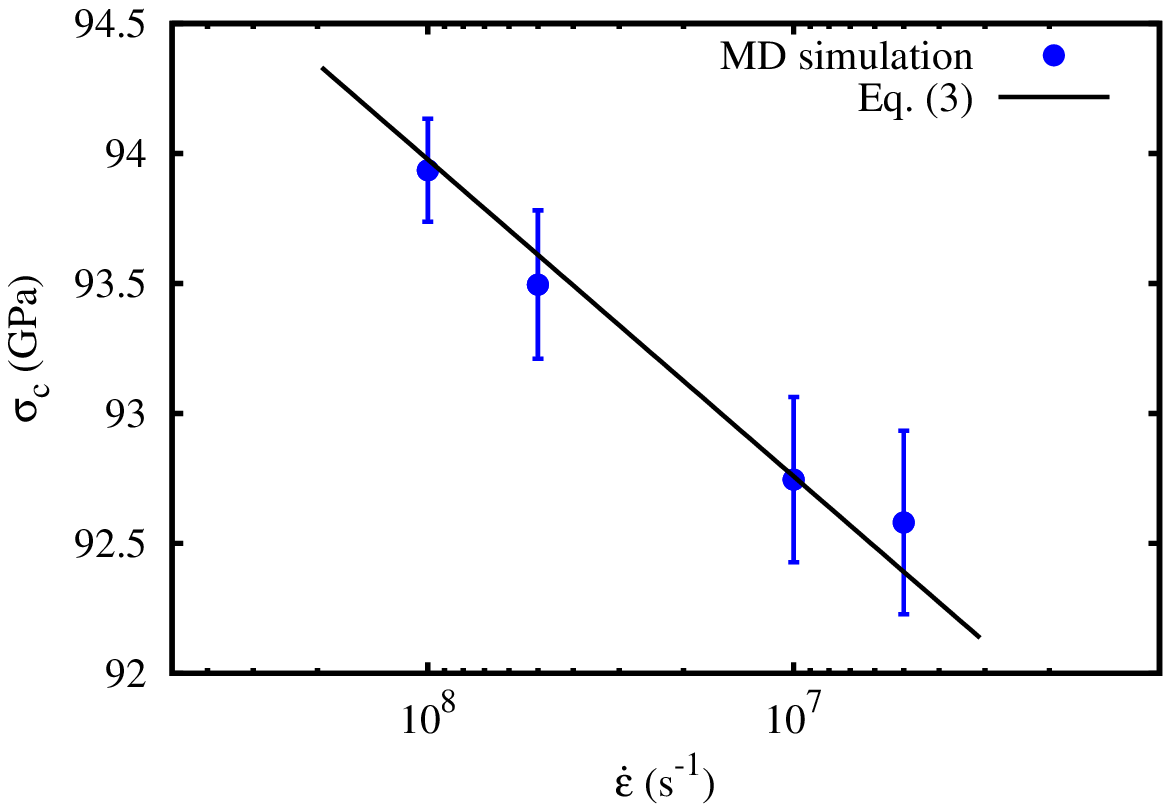}
\caption{Crack formation stresses $\sigma_\text{c}$ for pristine graphene loaded along the
armchair direction at T = 300 K.  Eq. (3) from the main text is overlaid with
$\text{E}_\text{o}$ = 4.93 eV and shows excellent agreement with MD data.  Error
bars are reported as 95\% confidence (n = 10).}
\label{sup2}
\end{figure}

In order to validate the energy barrier formulation used in this study, uniaxial
tensile simulations of pristine graphene are performed in the armchair loading
orientation.  Figure \ref{sup2} presents the variation in crack formation stress
with respect to strain rate at 300 K in armchair graphene using the MD
methodology outlined in the main text.  As expected, MD simulations predict a
monotonic decrease in $\sigma_\text{c}$ with respect to strain rate.  Eq.
(3) of the main text is overlaid with the collected MD data and shows excellent
agreement with simulation results.  The relevant parameters used in Eq.
(3) are $\text{E}_\text{o}$ = 4.93 eV \cite{Brenner2002},
$\tau_\text{o}$ = 0.1 ps \cite{Zhurkov1965}, and $\text{n}_\text{s}$ = 11808
=1.5N, where 1.5N is the number of bonds in an N atom system.  An activation
volume of 7.78 \AA$^\text{3}$ is obtained from fitting.  The activation volume
is approximately 1/2 the volume of the graphene unit cell and is reasonable for
brittle fracture.
\clearpage

\subsection*{Comparative energy barrier analysis at T = 450 K}
\begin{figure}[!h]
\centering
\includegraphics[width=0.8\textwidth, keepaspectratio]{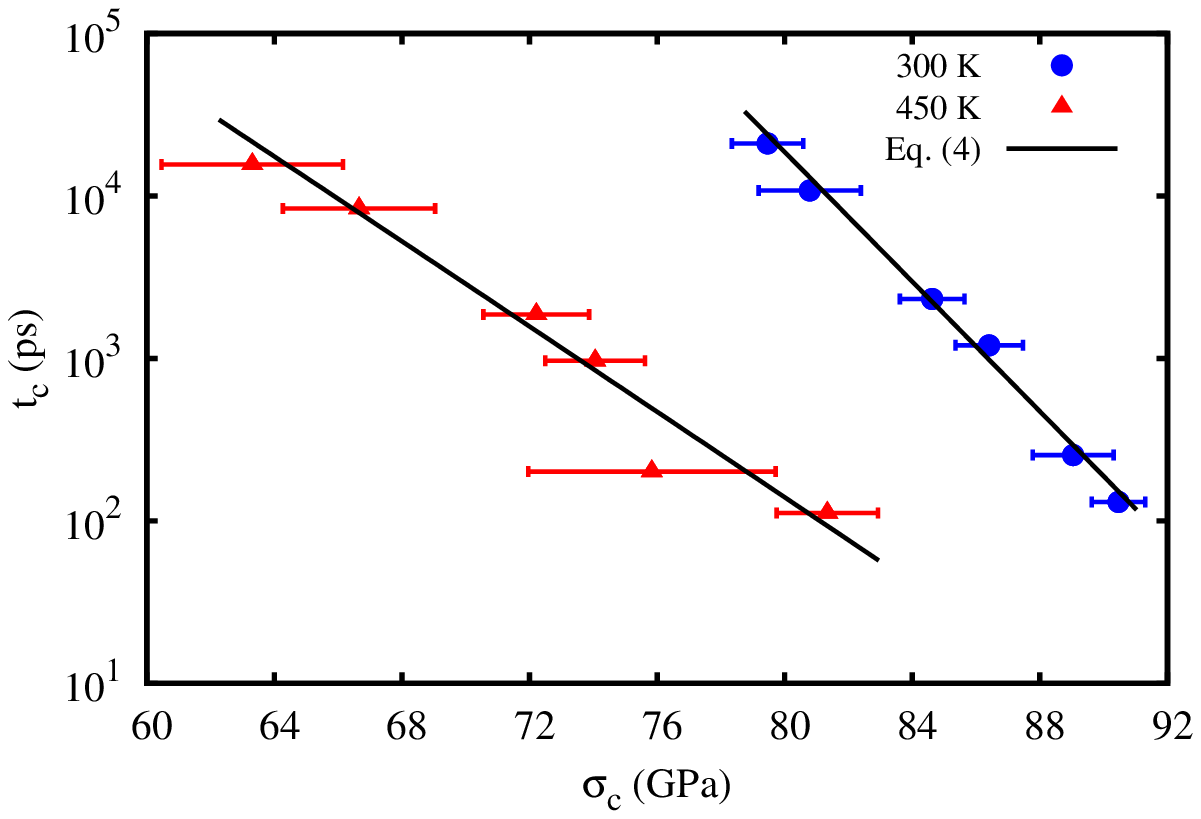}
\caption{Expectation times t$_\text{c}$ of the $\Sigma$ 7 tilt boundary at T = 300 and 450
K.  The activation energy and volume parameters are obtained from fitting of data in
Figure 7 and are used here with Eq. (4) from the main text, showing excellent
agreement with MD data.  Error bars are reported as 95\% confidence (n = 10).}
\label{sup3}
\end{figure}

As a further validation step, a parallel energy barrier analysis of the $\Sigma$
7 tilt boundary is conducted at T = 450 K.  The results of the expectation time
data collected from MD studies are presented in Figure \ref{sup3}.  Using the
activation energies and fitting parameters obtained from the T= 300 K dataset,
plotting of Eq. (4) from the main text shows excellent agreement at T= 450 K,
highlighting the robustness of the energy barrier formulation.  

\bibliography{refs}